\documentclass[aps,prb,onecolumn,groupedaddress]{revtex4}

\usepackage[english]{babel}

\usepackage{graphicx}
\usepackage[caption=false]{subfig}
\usepackage{dcolumn}

\usepackage{bm}
\usepackage{epsf}
\usepackage{color,ulem}
\usepackage{units}

\begin{document}

\title{Antibiotic resistance: a physicist's view}
\author{Rosalind Allen and Bart{\l}omiej Waclaw}
\affiliation{SUPA, School of Physics and Astronomy, The University of Edinburgh, Mayfield Road, Edinburgh EH9 3JZ, United Kingdom\\
and\\
Centre for Synthetic and Systems Biology, The University of Edinburgh}

\begin{abstract}
The problem of antibiotic resistance poses challenges across many disciplines. One such challenge is to understand the fundamental science of how antibiotics work, and how resistance to them can emerge. This is an area where physicists can make important contributions. Here, we highlight cases where this is already happening, and suggest directions for  further physics involvement in antimicrobial research.
\end{abstract}

\maketitle

The emergence and spread of bacterial infections that are resistant to antibiotic treatment, combined with the lack of development of new antibiotics, pose a global health problem that is now well recognised \cite{organization_who_2014,organization_who_2001,correspondent_antibiotics_2001,us_antibiotic_2013,uk_uk_2013}. Tackling antimicrobial resistance (AMR) requires coordinated, cross-disciplinary effort: relevant themes include clinical medicine, microbiology, diagnostics, drug discovery, epidemiology, evolutionary biology, global public health policy, veterinary science and agriculture. What role can physicists play in this spectrum of activities? Some of these themes obviously require physics tools, for example to aid the development of new experimental methods and devices. We argue here that a `physics-inspired' approach to basic science also has a prominent role to play in the effort to tackle AMR. 
\\

For a new, antibiotic-resistant, infectious bacterial strain to become a clinical problem, three events must occur. First, an individual pathogenic bacterium must acquire resistance to the antibiotic in question. This could happen via a spontaneous mutation in one of its genes, which might for example  render a target protein less susceptible to the antibiotic by modification of the antibiotic binding site. Alternatively, the pathogenic bacterium could gain a  gene encoding antibiotic resistance via horizontal transfer of DNA from a different bacterial strain. Second, the newly resistant bacterium must proliferate such that its resistance-encoding gene spreads in the local bacterial population and  cannot be wiped out through random fluctuations in the number of organisms carrying this gene. Third, the resistant strain must spread beyond the local bacterial population where it originated, until it infects a significant number of humans and becomes clinically relevant. These events occur on widely varying length and time scales, from those of molecules (e.g. a mutational event in a DNA strand) to those of macroscopic objects (bacterial biofilms, host animals, or even whole ecosystems), and they involve processes that relate directly to the realms of soft matter, chemical and statistical physics. \\

On the molecular level, physical scientists are already contributing to our understanding of how  antibiotics bind to their cellular targets, using both computer simulations and novel imaging techniques  \cite{tieleman_amp_2001,ndieyira_nanomechanical_2008,shaw_multidimensional_2015}. At the level of a bacterial cell, questions arise as to whether an antibiotic kills, or inhibits, a bacterial cell, via direct inhibition of its target (e.g. the cell wall synthesis machinery for beta-lactam antibiotics or the protein synthesis machinery for macrolide antibiotics), or via other, downstream effects \cite{kohanski_2007,Palmer_2014,Wei_2011}. Here, physicists can contribute by developing simple models for how the complex network of reactions that constitutes bacterial physiology response to the antibiotic-induced stress \cite{Greulich_2015,Weisse_2015}. At the level of a bacterial population, physical interactions between cells and their environment shape the self-assembly of spatially-structured bacterial conglomerates such as biofilms that form on medical implants \cite{donlan_2002}. From a physics point of view, the interplay between biological phenomena such as growth and physical phenomena such as chemical  diffusion and physical forces provides many interesting questions. For example, biofilms are often surrounded by a secreted polymer matrix whose physical properties (e.g. viscosity) may affect how the biofilm assembles and how it responds to drug treatment \cite{flemming_2010}. Moreoever nutrient and drug  gradients  can emerge in biofilms due to the interplay between growth and chemical transport; these can affect biofilm structure \cite{dockery_2001,melaugh_2016,kragh_2016} and potentially also the rate of evolution of resistant bacteria \cite{zhang_acceleration_2011,hermsen_rapidity_2012,greulich_mutational_2012}. Other population-level phenomena of interest to physicists include stochastic differences in the behaviour of individual cells, caused by noise in gene expression \cite{raser_2005}, which can have drastic consequences for the response of the population to antibiotic treatment \cite{balaban_2004}.\\ 

``A physics-like'' approach thus has a role to play in many aspects of AMR if we define such approach as a belief that biological
processes can be explained by a combination of simple, yet quantitative experiments and mathematical modelling. In the remainder of this article, we highlight three areas where such physics-like approaches are already proving successful, and we also comment on promising directions for future research. \\

\begin{figure}
	\centering
		\includegraphics[width=14cm]{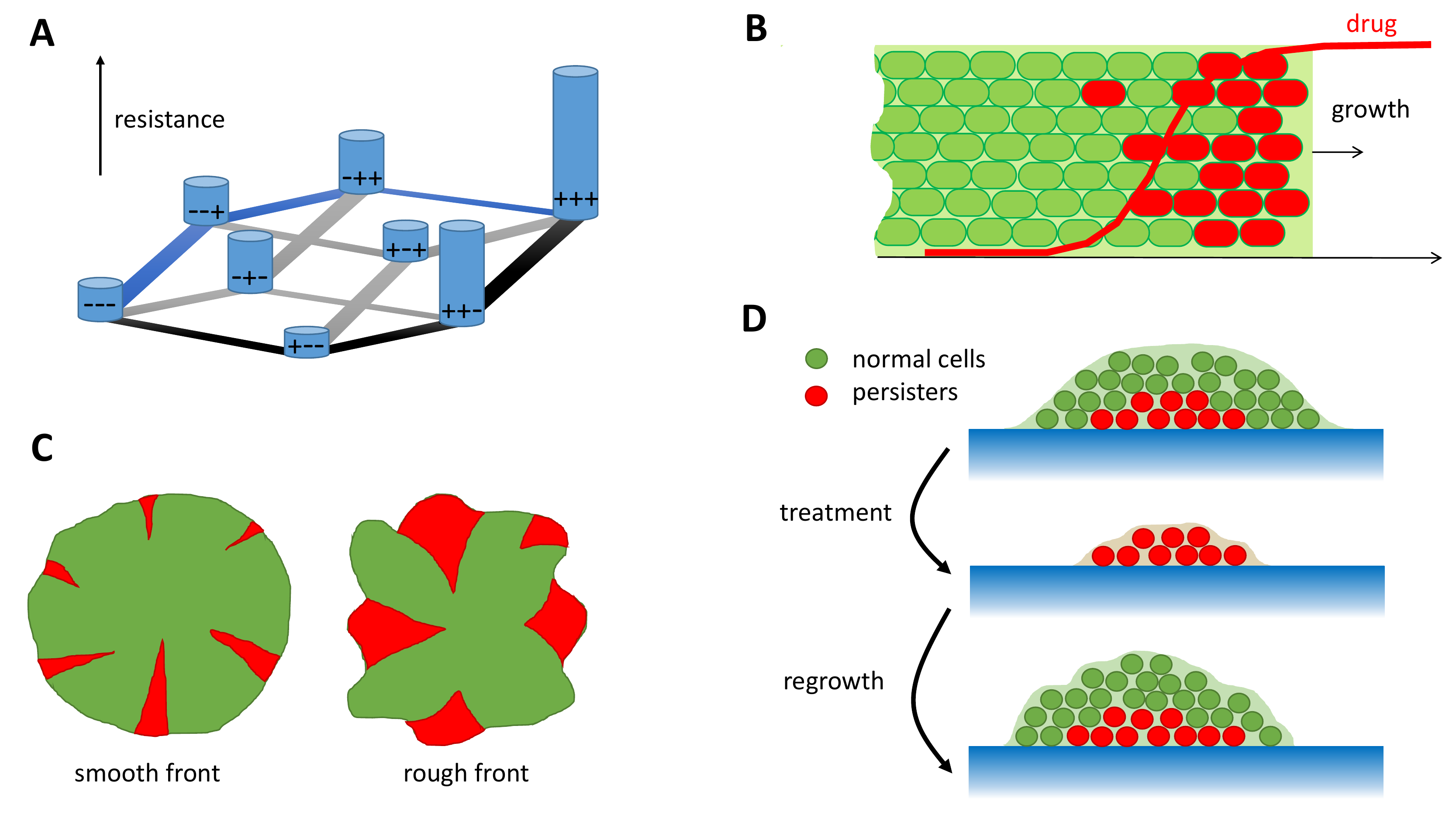}
	\caption{(A) Example of a resistance landscape with three mutations. ``+''/``-'' denote the presence/absence of a particular mutation, hence the sensitive strain is represented by $---$ and the most-resistant mutant strain by $+++$. The level of resistance is indicated by the height of the bars, and permitted mutations are marked by gray lines. The mutant $+++$ can be reached by many paths, for example $(---)\to(--+)\to(-++)\to(+++)$ along which resistance increases monotonously (blue lines), or $(---)\to(+--)\to(++-)\to(+++)$ which has a ``valley'' at $+--$ (black lines). (B) Evolution in the presence of an antibiotic gradient for an expanding population. The fate of mutant red cells depends on whether they arise in the ``bulk'' or at the front of the population wave. In the first case there is no selective pressure due to low drug concentration and the mutant does not spread. However, if the mutant arises at the front, it benefits from the access to nutrients/higher growth rate, and it spreads quickly. (C) If resistant cells arise in microbial colonies, they form spatial ``sectors''. The number and the size of such sectors depends on the roughness of the colony's frontier, which in turn is affected by physical interactions between the cells and their environment. (D) A microbial population often contains ``persister'' cells which grow slowly but are resistant to drugs.	If the population is treated with antibiotics, persisters can survive the treatment, switch back to the growing state and cause regrowth.}
\end{figure}

\noindent{\bf{Pathways to resistance}}\\

An active area of current research focuses on how antibiotic resistance evolves ``de novo'', i.e. by genetic mutation in bacterial strains that are not initially resistant (as opposed to via gene transfer from an already resistant strain). Typically, the process of resistance evolution involves not just one genetic mutation but a sequence of mutations. This mutational ``pathway to resistance'' is one of many possible sequences of mutations in a hugely multidimensional space made up of all the possible genetic variants (genotypes) of the organism (Fig. 1A). To understand how resistance evolves we must therefore understand the structure of this ``resistance landscape''.

Two alternative models represent extreme limits of the resistance landscape. In the first model, the level of resistance changes randomly with each mutation (known as a ``maximally rugged'', or ``House of Cards'' landscape \cite{kingman_1978}), while in the second,  mutations are additive, such that the total resistance is the sum of the contributions of the various mutations that a bacterium has acquired (this corresponds to a smooth fitness landscape). Using a statistical physics approach, an analysis of measured resistance landscapes for several different antibiotics suggests that these landscapes are neither fully random nor fully additive, but that they can be approximated by a ``rough Mount Fuji'' model, which essentially  corresponds to a superposition of the rugged and smooth models \cite{Szendro_quantitative_2013,de_visser_empirical_2014}. Further computer modelling \cite{szendro_predictability_2013} has shown that the pathways followed by evolution on such landscapes are most predictable in intermediate-size populations ($N\mu^2\ll 1 \ll N\mu$, where $N$ is the population size and $\mu$ is the mutation probability) whereas the evolution of resistance is predicted to be less reproducible in either very small or very large populations. 

The predictability of the evolution of resistance has recently  been tested experimentally by growing bacteria in antibiotic under constant selective pressure, using a ``morbidostat'' \cite{toprak_evolutionary_2012,toprak_building_2013}. This is a device in which the population density and growth rate of a bacterial population are constantly monitored, and the dosage of antibiotic to which it is subjected is adjusted to maintain (on average) a fixed growth rate. Thus, as the population evolves resistance, the antibiotic dosage increases. By sequencing the DNA of bacteria sampled from the morbidostat as the population evolves, qualitative differences in the pathways by which resistance to different drugs emerges have been revealed. In particular, the evolution of resistance to the antibiotic trimethoprim has been found to be rather reproducible, occurring by a well-defined and universal sequence of mutations, but the evolution of resistance to the antibiotics chloramphenicol and doxycycline was found to show a less predictable pattern, occurring via different sequences of mutations in replicate experiments \cite{toprak_evolutionary_2012}.

Another ``simple yet smart'' experimental design has led to a different insight into the pathways leading to antibiotic resistance evolution. Here, bacteria are subjected to repeated brief exposures to antibiotic, alternating with periods of growth without antibiotic \cite{balaban_2014}.  Remarkably the bacteria become tolerant to the antibiotic by matching the duration of their ``lag time'', or the dormancy period before they start to grow when exposed to fresh nutrient medium, to the duration of antibiotic exposure \cite{balaban_2014}. This is an important finding because it suggests that tolerance to antibiotic exposure can  be achieved much more easily than full resistance, by tinkering with the cell's existing gene regulatory components, and that this may be a first step in the pathway to full resistance. Thus, to fully understand the pathways by which bacteria become resistant to antibiotics, we may need to study not just genetic mutations but also phenotypic changes that arise via gene regulation \cite{riesenfeld_adaptive_1997,adam_epigenetic_2008,breidenstein_pseudomonas_2011}.\\

\noindent{\bf{Evolution in spatially structured environments}}\\

Most studies of antibiotic resistance evolution assume that it happens in a well-mixed, spatially homogeneous environment. However, in infections, as well as more widely in the natural environment, bacterial populations are often highly spatially structured. For example, in biofilm infections bacteria deep within the biofilm are likely to experience quite different local concentrations of nutrients and, possibly, of antibiotic, compared with bacteria on the outside of the biofilm. Indeed, confocal microscopy images of lab-grown biofilms exposed to antibiotics show that some antibiotics act selectively on cells on the outside of the biofilm, while others selectively kill those on the inside \cite{pamp_2008}. Could the spatial gradients of antibiotic that arise during treatment of infections influence the evolution of antibiotic resistance?

Experimental and theoretical physicists have recently addressed this topic by investigating the growth and evolution of bacterial populations in the presence of controlled spatial gradients of antibiotic. In particular, Zhang {\it{et al}} constructed a microfluidic device which allowed antibiotic gradients to be imposed across a spatially expanding bacterial population \cite{zhang_acceleration_2011}. In this device, the evolution of {\it{E. coli}} bacteria which were resistant to the antibiotic ciprofloxacin occurred much more rapidly in the presence of an antibiotic gradient than in a uniform concentration of antibiotic. This work inspired the development of several theoretical models in which an expanding bacterial population invades a one-dimensional habitat containing a gradient of antibiotic \cite{greulich_mutational_2012,hermsen_rapidity_2012} (Fig. 1B). These models suggest  that the presence of an antibiotic gradient can accelerate the evolution of resistance because mutants that emerge at the tip of the expanding population wave are exposed to high antibiotic concentrations, for which they have a strong selective advantage, and because of the low population density at the tip of the population wave, they do not have to compete with less resistant genotypes. However this is not always the case; too steep a gradient can actually slow down the evolution of resistance, and the process also depends on the rate of bacterial migration \cite{hermsen_rapidity_2012} and the mutational pathway to resistance \cite{greulich_mutational_2012}. 

Other important recent work shows that, even in the absence of an antibiotic gradient, evolution can work very differently in a spatially structured population compared to a spatially well-mixed one. Using colonies of bacteria and yeast cells growing on the surface of a semi-solid agar gel, Hallatschek {\it et al}, and others, showed that population fluctuations at the front of an expanding microbial colony can lead to some cell lineages randomly spreading in the population while others die out \cite{hallatschek_2007,hallatschek_2010,korolev_2012,mitri_2015}. In these experiments, spreading lineages are strikingly visualised as sectors of differently-coloured fluorescent cells (Fig. 1C). Further work investigating the probability of fixation of mutants in these expanding populations suggests that the roughness of the growing front plays a crucial role \cite{gralka}. However, much more work is needed before we can translate these principles into a full understanding of how the spatial structure of a bacterial population, for example in a biofilm infection, links to its propensity to generate and fix harmful antibiotic-resistant mutants.\\

\noindent{\bf{Response of individual cells to antibiotics}}\\

Over the past decades, physicists have played a leading role in establishing the existence, and importance, of variation between individual cells within bacterial populations \cite{elowitz_2002,raj_2008,ackermann_2015}. This phenotypic heterogeneity arises from stochasticity in the molecular processes involved in gene expression, modulated by the networks of interactions between proteins and genes that regulate gene expression. From the perspective of antibiotic resistance, a particularly important example of phenotypic heterogeneity is the existence  within  bacterial populations of a subpopulation of ``persister cells'' which can survive antibiotic treatment (Fig. 1D). Classic experiments using microfluidic devices by Balaban {\it et al} visualised {\it{E. coli}} cells  switching into and out of the non-growing persister state \cite{balaban_2004}; yet the molecular mechanisms controlling this switch are still the topic of active research \cite{vega_signalling_2012,maisonneuve_2013}. 

Evidence is also now emerging that bacterial population heterogeneity in the response to antibiotics may be a more general phenomenon. Several different theoretical models for the effects of antibiotics on bacterial growth predict the existence of bistable regimes, in which some cells in the population show fast growth while others grow only very slowly, if at all \cite{elf_2006,Greulich_2015}. The models suggest that this growth bistability can have a number of origins, including the interplay between bacterial growth and drug dilution \cite{elf_2006} and an irreversible ``annihilation'' reaction between an antibiotic and its target  \cite{Greulich_2015}. Joint experimental and theoretical work has also shown that  growth bistability can arise from the partitioning of cellular resources between growth and production of proteins leading to antibiotic resistance \cite{deris_2013}. Whether, and how, these bistable growth regimes are related to the persister phenomenon, remains unclear. From the clinical point of view, we would also like to understand which drugs produce a heterogeneous response and how this impacts both on treatment strategies and on the evolution of resistance.

From an experimental point of view, investigations of bacterial phenotypic heterogeneity often rely on microfluidic technology which combines physical, chemical, and engineering knowledge to image and manipulate individual bacterial cells. Devices such as the ``mother machine'' \cite{wang_2010} allow the proliferation of individual cells to be tracked under constant conditions over long times in the microscope, and these devices are beginning to reveal interesting information about the response to antibiotics \cite{lambert_2015}. Another recently-developed microfluidic technique,  Microscopy Assisted Cell Screening, provides a way to rapidly image thousands of cells immediately after sampling them from a growth device \cite{okumus_2016}. This approach can be used, for example, to measure population heterogeneity in processes like DNA damage, which involve small numbers of molecules per cell. 
\\

\noindent{\bf{Perspective}}\\

The topics discussed above highlight important contributions that are being made by physicists to our understanding of antibiotic action and to AMR. We believe that this is just the beginning. Many questions remain to be addressed about how bacteria respond (and become resistant) to antibiotics, and physicists have an important role to play in this effort. As a first example, it is imperative to gain better understanding of how bacterial cells interact mechanically with one another and with their environment. Mechanical interactions appear to be very important in bacterial self-assembly \cite{volfson_biomechanical_2008,asally_cover:_2012,farrell_mechanically_2013,grant_2014}, yet our limited knowledge of these interactions prevents us from building accurate models of how spatially structured infections like bacterial biofilms form. As a second example, horizontal gene transfer -- the transmission of genes encoding antibiotic resistance between (potentially) unrelated bacteria by direct transfer of DNA -- has been little studied in a ``physics'' context \cite{venegas-ortiz_speed_2014,court_parasites_2013}, yet it is very important in clinically relevant antibiotic-resistant infections. It would be very interesting to investigate how physical factors such as the forces existing between adjacent bacterial cells in a colony or biofilm affect the rate of gene transfer \cite{kuba+BW}.

We also believe that AMR can provide a rich source of more abstract problems in areas of physics from non-equilibrium statistical mechanics to soft matter physics and fluid mechanics. For example, the growth of bacterial biofilms shows a non-equilibrium phase transition (or fingering instability) between rough and smooth modes of interface growth \cite{dockery_2001}. Other examples include random walk models inspired by the dynamics of mutant sectors in expanding bacterial populations \cite{hallatschek_2007,hallatschek_2010}, and stochastic differential equation models to describe the emergence of waves of resistant mutants in an antibiotic gradient \cite{greulich_mutational_2012}. Physics models that are inspired by AMR may also be transferable to other biological fields - for example the evolution of bacterial antibiotic resistance has important analogies with the evolution of drug resistance in cancer tumours \cite{lambert_2011}, while models for the dynamics of bacterial and viral infections also have many similarities. 

Finally, we believe that an important goal of future research in AMR, by physicists and others, must be to make a link between the findings of simple laboratory experiments and theoretical models, and what happens in real infections, in clinical settings \cite{BW_book_chapter}. Existing areas of progress in this direction include using observations on how antibiotics interact with the physiology of bacterial cells to suggest more effective clinical treatment strategies \cite{brynildsen_2013,allison_2011,Greulich_2015}, using systematic measurements of thousands of bacterial growth curves, combined with simple theoretical models, to predict the clinical effectiveness of multi-drug therapies \cite{bollenbach_2015,bollenbach_2009}, and tracking mutational pathways in {\it{in vivo}} infections \cite{lieberman_2014}. Future challenges will include understanding the role of spatially structured infections in the evolution of clinical antibiotic resistance and the impact of bacterial population heterogeneity on clinical responses to treatment.\\

\noindent{\bf{Conclusion}}\\

Achieving a better understanding of how antibiotic resistance emerges and spreads could help us to design strategies to prevent this from happening, for both current and future antibiotics. This is a goal to which physicists can contribute, not just by developing new machines and software tools, but also by designing simple yet insightful experimental and theoretical models to test basic principles. This work should not be carried out in isolation, but in coordination with the broad spectrum of other efforts that are being made to tackle the problem of antibiotic resistance. To make this possible, however, support is needed in the form of funding. This requires a breakdown of traditional discipline barriers. Indeed, for the breakthroughs highlighted in this article it is not relevant to ask ``is it physics''? or ``is it biology'', but simply ``is it ground-breaking science''?  It also requires a breakdown of national barriers, to allow the most talented scientists, within what is still a rather small international field, to work productively together without restrictions imposed by funding regulations. 
\\

\noindent{\bf Acknowledgements}\\

This article was inspired by a Royal Society International Scientific Seminar on ``Antimicrobial resistance: how can physicists help?'', which took place at Chicheley Hall, UK, on October 28th and 29th 2015.  RJA is supported by a Royal Society University Research Fellowship and by ERC Consolidator Grant 682237-`EVOSTRUC'. BW is supported by a Royal Society of Edinburgh Research Fellowship.


\begin{thebibliography}{10}

\bibitem{organization_who_2014}
World~Health Organization.
\newblock Antimicrobial resistance: {G}lobal report on surveillance.
\newblock 2014.

\bibitem{organization_who_2001}
World~Health Organization and {others}.
\newblock {WHO} global strategy for containment of antimicrobial resistance.
\newblock 2001.

\bibitem{correspondent_antibiotics_2001}
Fergus Walsh~Medical correspondent.
\newblock Antibiotics resistance 'as big a risk as terrorism' - medical chief,
  2001.

\bibitem{us_antibiotic_2013}
Centres for Disease~Control and Prevention (US).
\newblock Antibiotic resistance threats in the {United} {States}.
\newblock 2013.

\bibitem{uk_uk_2013}
SC~Davies and N~Gibbens.
\newblock Uk five year antimicrobial resistance strategy 2013--2018.
\newblock {\em London: Department of Health}, 2013.

\bibitem{tieleman_amp_2001}
D.~P. Tieleman and M.~S.~P. Sansom.
\newblock Molecular dynamics simulations of antimicrobial peptides: {F}rom
  membrane binding to trans-membrane channels.
\newblock {\em International Journal of Quantum Chemistry}, 83:166--179, 2001.

\bibitem{ndieyira_nanomechanical_2008}
J.~W. Ndieyira, M.~Watari, A.~D. Barrera, D.~Zhou, M.~V{\"{g}}tli,
  M.~Batchelor, M.~A. Cooper, T.~Strunz, M.~A. Horton, C.~Abell, T.~Rayment,
  G.~Aeppli, and R.~A. Mc{K}endry.
\newblock Nanomechanical detection of antibiotic –- mucopeptide binding in a
  model for superbug drug resistance.
\newblock {\em Nature Nanotechnology}, 3(11):691--696, 2008.

\bibitem{shaw_multidimensional_2015}
D.~J. Shaw, K.~Adamczyk, P.~W. J.~M. Frederix, N.~Simpson, K.~Robb, G.~M.
  Greetham, M.~Towrie, A.~W. Parker, P.~A. Hoskisson, and N.~T. Hunt.
\newblock Multidimensional infrared spectroscopy reveals the vibrational and
  solvation dynamics of isoniazid.
\newblock {\em The Journal of Chemical Physics}, 142(21):212401, 2015.

\bibitem{kohanski_2007}
M.~A. Kohanski, D.~J. Dwyer, B.~Hayete, C.~A. Lawrence, and J.~J. Collins.
\newblock A common mechanism of cellular death induced by bactericidal
  antibiotics.
\newblock {\em Cell}, 130:797--810, 2007.

\bibitem{Palmer_2014}
A.~C. Palmer and R.~Kishony.
\newblock Opposing effects of target overexpression reveal drug mechanisms.
\newblock {\em Nature Communications}, 5:4296, 2014.

\bibitem{Wei_2011}
J.-R. Wei, V.~Krishnamoorthy, K.~Murphy, J.-H. Kim, D.~Schnappinger, T.~Alber,
  C.~M. Sassetti, K.~Y. Rhee, and E.~J. Rubin.
\newblock Depletion of antibiotic targets has widely varying effects on growth.
\newblock {\em Proc. Natl. Acad. Sci. {USA}}, 108:4176--4181, 2011.

\bibitem{Greulich_2015}
P.~Greulich, M.~Scott, M.~R. Evans, and R.~J. Allen.
\newblock Growth-dependent bacterial susceptibility to ribosome-targeting
  antibiotics.
\newblock {\em Molecular Systems Biology}, 11:796, 2015.

\bibitem{Weisse_2015}
A.~Y. Weisse, D.~A. Oyarzun, V.~Danos, and P.~S. Swain.
\newblock Mechanistic links between cellular trade-offs, gene expression, and
  growth.
\newblock {\em Proc. Natl. Acad. Sci. {USA}}, 112:E1038, 2015.

\bibitem{donlan_2002}
R.~M. Donlan.
\newblock Biofilms: {M}icrobial life on surfaces.
\newblock {\em Emerg. Infect. Dis.}, 8:881--890, 2002.

\bibitem{flemming_2010}
H.-C. Flemming and J.~Wingender.
\newblock The biofilm matrix.
\newblock {\em Nature Reviews Microbiology}, 8:623--633, 2010.

\bibitem{dockery_2001}
J.~Dockery and I.~Kapper.
\newblock Finger formation in biofilm layers.
\newblock {\em {SIAM} J. Appl. Math.}, 62:853--869, 2001.

\bibitem{melaugh_2016}
G.~Melaugh, J.~Hutchison, K.~N. Kragh, Y.~Irie, A.~Roberts, T.~Bjarnsholt,
  S.~P. Diggle, V.~D. Gordon, and R.~J. Allen.
\newblock Shaping the growth behaviour of biofilms initiated from bacterial
  aggregates.
\newblock {\em {PL}o{S} {ONE}}, 11:e0149683, 2016.

\bibitem{kragh_2016}
K.~N. Kragh, J.~Hutchison, G.~Melaugh, C.~Rodesney, A.~E.~L. Roberts, Y.~Irie,
  P.~O. Jensen, S.~P. Diggle, R.~J. Allen, V.~D. Gordon, and T.~Bjarnsholt.
\newblock Role of multicellular aggregates in biofilm formation.
\newblock {\em m{B}io}, 7:e00237--16, 2016.

\bibitem{zhang_acceleration_2011}
Q.~Zhang, G.~Lambert, D.~Liao, H.~Kim, K.~Robin, C.-{K}. Tung, N.~Pourmand, and
  R.~H. Austin.
\newblock Acceleration of {Emergence} of {Bacterial} {Antibiotic} {Resistance}
  in {Connected} {Microenvironments}.
\newblock {\em Science}, 333(6050):1764--1767, 2011.

\bibitem{hermsen_rapidity_2012}
R.~Hermsen, J.~B Deris, and T.~Hwa.
\newblock On the rapidity of antibiotic resistance evolution facilitated by a
  concentration gradient.
\newblock {\em Proceedings of the National Academy of Sciences},
  109(27):10775--10780, 2012.

\bibitem{greulich_mutational_2012}
P.~Greulich, B.~Waclaw, and R.~J. Allen.
\newblock Mutational {pathway} {determines} {whether} {drug} {gradients}
  {accelerate} {evolution} of {drug}-{resistant} {cells}.
\newblock {\em Physical Review Letters}, 109(8), 2012.

\bibitem{raser_2005}
J.~M. Raser and E.~K. O'{S}hea.
\newblock Noise in gene expression: origins, consequences, and control.
\newblock {\em Science}, 309:2010--2013, 2005.

\bibitem{balaban_2004}
N.~Q. Balaban, J.~Merrin, R.~Chait, L.~Kowalik, and S.~Leibler.
\newblock Bacterial persistence as a phenotypic switch.
\newblock {\em Science}, 305:1622--1625, 2004.

\bibitem{kingman_1978}
J.~F.~C. Kingman.
\newblock A simple model for the balance between selection and mutation.
\newblock {\em J. Appl. Prob.}, 15:1--12, 1978.

\bibitem{Szendro_quantitative_2013}
I.~G. Szendro, M.~F. Schenk, J.~Franke, J.~Krug, and J.~A. G. M.~de Visser.
\newblock Quantitative analyses of empirical fitness landscapes.
\newblock {\em Journal of Statistical Mechanics: Theory and Experiment},
  2013(01):P01005, 2013.

\bibitem{de_visser_empirical_2014}
J.~A. G.~M. de~Visser and J.~Krug.
\newblock Empirical fitness landscapes and the predictability of evolution.
\newblock {\em Nature Reviews Genetics}, 15(7):480--490, 2014.

\bibitem{szendro_predictability_2013}
I.~G. Szendro, J.~Franke, J.~A. G. M.~de Visser, and J.~Krug.
\newblock Predictability of evolution depends nonmonotonically on population
  size.
\newblock {\em Proceedings of the National Academy of Sciences},
  110(2):571--576, 2013.

\bibitem{toprak_evolutionary_2012}
E.~Toprak, A.~Veres, J.-B. Michel, R.~Chait, D.~L Hartl, and R.~Kishony.
\newblock Evolutionary paths to antibiotic resistance under dynamically
  sustained drug selection.
\newblock {\em Nat Genet}, 44(1):101--105, 2012.

\bibitem{toprak_building_2013}
E.~Toprak, A.~Veres, S.~Yildiz, J.~M. Pedraza, R.~Chait, J.~Paulsson, and
  R.~Kishony.
\newblock Building a morbidostat: an automated continuous-culture device for
  studying bacterial drug resistance under dynamically sustained drug
  inhibition.
\newblock {\em Nature Protocols}, 8(3):555--567, 2013.

\bibitem{balaban_2014}
O.~Fridman, A.~Goldberg, I.~Ronin, N.~Shoresh, and N.~Q. Balaban.
\newblock Optimization of lag time underlies antibiotic tolerance in evolved
  bacterial populations.
\newblock {\em Nature}, 513:418--421, 2014.

\bibitem{riesenfeld_adaptive_1997}
C.~Riesenfeld, M.~Everett, L.~J. Piddock, and B.~G. Hall.
\newblock Adaptive mutations produce resistance to ciprofloxacin.
\newblock {\em Antimicrobial Agents and Chemotherapy}, 41(9):2059--2060, 1997.

\bibitem{adam_epigenetic_2008}
M.~Adam, B.~Murali, N.~O. Glenn, and S.~S. Potter.
\newblock Epigenetic inheritance based evolution of antibiotic resistance in
  bacteria.
\newblock {\em BMC Evolutionary Biology}, 8(1):1, 2008.

\bibitem{breidenstein_pseudomonas_2011}
E.~B.~M. Breidenstein, C.~de~la Fuente-Núñez, and R.~E.~W. Hancock.
\newblock Pseudomonas aeruginosa: all roads lead to resistance.
\newblock {\em Trends in Microbiology}, 19(8):419--426, 2011.

\bibitem{pamp_2008}
S.~J. Pamp, M.~Gjermansen, H.~K. Johansen, and T.~Tolker-{N}ielsen.
\newblock Tolerance to the antimicrobial peptide colistin in {\it{{p}seudomonas
  aeruginosa}} biofilms is linked to metabolically active cells, and depends on
  the {\it{pmr}} and {\it{mex{ab}-opr{m}}} genes.
\newblock {\em Molecular Microbiology}, 68:223--240, 2008.

\bibitem{hallatschek_2007}
O.~Hallatschek, P.~Hersen, S.~Ramanathan, and D.~R. Nelson.
\newblock Genetic drift at expanding frontiers promotes gene segregation.
\newblock {\em Proc. Natl. Acad. Sci. {USA}}, 104:19926--19930, 2007.

\bibitem{hallatschek_2010}
O.~Hallatschek and D.~R. Nelson.
\newblock Life at the front of an expanding population.
\newblock {\em Evolution}, 64:193--206, 2010.

\bibitem{korolev_2012}
K.~S. Korolev, M.~J.~I. M{\"{u}}ller, N.~Karahan, A.~W. Murray, O.~Hallatschek,
  and D.~R. Nelson.
\newblock Selective sweeps in growing microbial colonies.
\newblock {\em Physical Biology}, 9:026008, 2012.

\bibitem{mitri_2015}
S.~Mitri, E.~Clarke, and K.~R. Foster.
\newblock Resource limitation drives spatial organization in microbial groups.
\newblock {\em {ISME} Journal}, page 2015.208, 2015.

\bibitem{gralka}
M.~Gralka, F.~Stiewe, F.~Farrell, W.~Moebius, B.~Waclaw, and O.~Hallatschek.
\newblock Allele surfing promotes microbial adaptation from standing variation.
\newblock {\em bio{R}xiv}, page 10.1101/049353, 2016.

\bibitem{elowitz_2002}
M.~B. Elowitz, A.~J. Levine, E.~D. Siggia, and P.~S. Swain.
\newblock Stochastic gene expression in a single cell.
\newblock {\em Science}, 297:1183--1186, 2002.

\bibitem{raj_2008}
A.~Raj and A.~Oudenaarden.
\newblock Stochastic gene expression and its consequences.
\newblock {\em Cell}, 135:216--226, 2008.

\bibitem{ackermann_2015}
M.~Ackermann.
\newblock A functional perspective on phenotypic heterogeneity in
  microorganisms.
\newblock {\em Nature Reviews Microbiology}, 13:497--508, 2015.

\bibitem{vega_signalling_2012}
N.~M. Vega, K.~R. Allison, A.~S. Khalil, and J.~J. Collins.
\newblock Signaling-mediated bacterial persister formation.
\newblock {\em Nature Chemical Biology}, 8:431--433, 2012.

\bibitem{maisonneuve_2013}
E.~Maisonneuve, M.~Castro-{C}amargo, and K.~Gerdes.
\newblock {(p)}pp{G}pp controls bacterial persistence by stochastic induction
  of toxin-antitoxin activity.
\newblock {\em Cell}, 154:1140--1150, 2013.

\bibitem{elf_2006}
J.~Elf, K.~Nilsson, T.~Tenson, and M.~Ehrenberg.
\newblock Bistable bacterial growth rate in response to antibiotics with low
  membrane permeability.
\newblock {\em Physical Review Letters}, 97:258104, 2006.

\bibitem{deris_2013}
J.~B. Deris, M.~Kim, Z.~Zhang, H.~Okano, R.~Hermsen, A.~Groisman, and T.~Hwa.
\newblock The innate growth bistability and fitness landscapes of antibiotic
  resistant bacteria.
\newblock {\em Science}, 342:1237435, 2012.

\bibitem{wang_2010}
P.~Wang, L.~Robert, J.~Pelletier, W.~L. Dang, F.~Taddei, A.~Wright, and S.~Jun.
\newblock Robust growth of {\it{{e}scherichia coli}}.
\newblock {\em Current Biology}, 20:1099--1103, 2010.

\bibitem{lambert_2015}
G.~Lambert and E.~Kussell.
\newblock Quantifying selective pressures driving bacterial evolution using
  lineage analysis.
\newblock {\em Physical Review X}, 5:011016, 2015.

\bibitem{okumus_2016}
B.~Okumus, D.~Landgraf, G.~C. Lai, S.~Bakhsi, J.~C. Arias-{C}astro, S.~Yildiz,
  D.~Huh, R.~Fernandez-{L}opez, C.~N. Peterson, E.~Toprak, M.~El-{K}aroui, and
  J.~Paulsson.
\newblock Mechanical slowing down of cytoplasmic diffusion allows in vivo
  counting of proteins in individual cells.
\newblock {\em Nature Communications}, (in press), 2016.

\bibitem{volfson_biomechanical_2008}
D.~Volfson, S.~Cookson, J.~Hasty, and L.S. Tsimring.
\newblock Biomechanical ordering of dense cell populations.
\newblock {\em Proceedings of the National Academy of Sciences}, 105(40):15346,
  2008.

\bibitem{asally_cover:_2012}
M.~Asally, M.~Kittisopikul, P.~Rue, Y.~Du, Z.~Hu, T.~Cagatay, A.~B. Robinson,
  H.~Lu, J.~Garcia-Ojalvo, and G.~M. Suel.
\newblock {Localized} cell death focuses mechanical forces during 3d patterning
  in a biofilm.
\newblock {\em Proceedings of the National Academy of Sciences},
  109(46):18891--18896, 2012.

\bibitem{farrell_mechanically_2013}
F.~D.~C. Farrell, O.~Hallatschek, D.~Marenduzzo, and B.~Waclaw.
\newblock Mechanically {Driven} {Growth} of {Quasi}-{Two}-{Dimensional}
  {Microbial} {Colonies}.
\newblock {\em Physical Review Letters}, 111(16), 2013.

\bibitem{grant_2014}
M.~A.~A. Grant, B.~Waclaw, R.~J. Allen, and P.~Cicuta.
\newblock The role of mechanical forces in the planar-to-bulk transition in
  growing {\it{{e}scherichia coli}} microcolonies.
\newblock {\em J. R. Soc. Interface}, 11:20140400, 2014.

\bibitem{venegas-ortiz_speed_2014}
J.~Venegas-Ortiz, R.~J. Allen, and M.~R. Evans.
\newblock Speed of {invasion} of an {expanding} {population} by a
  {horizontally} {transmitted} {trait}.
\newblock {\em Genetics}, 196(2):497--507, 2014.

\bibitem{court_parasites_2013}
S.~J. Court, R.~A. Blythe, and R.~J. Allen.
\newblock Parasites on parasites: {Coupled} fluctuations in stacked contact
  processes.
\newblock {\em EPL (Europhysics Letters)}, 101(5):50001, 2013.

\bibitem{kuba+BW}
K.~Pastuszak and B.~Waclaw.
\newblock A physics-explicit model of plasmid conjugation in an expanding
  bacterial colony.
\newblock {\em in preparation}.

\bibitem{lambert_2011}
G.~Lambert, L.~Estevez-{S}almeron, S.~Oh, D.~Liao, B.~M. Emerson, T.~D. Tlsty,
  and R.~H. Austin.
\newblock An analogy between the evolution of drug resistance in bacterial
  communities and malignant tissues.
\newblock {\em Nature Reviews Cancer}, 11:375--382, 2011.

\bibitem{BW_book_chapter}
B.~Waclaw.
\newblock {\em Evolution of drug resistance in bacteria, in Biophysics of
  Infection, edited by M. Leake}.
\newblock Springer, 2016.

\bibitem{brynildsen_2013}
M.~P. Brynildsen, J.~A. Winkler, C.~S. Spina, C.~I. Mac{D}onald, and J.~J.
  Collins.
\newblock Potentiating antibacterial activity by predictably enhancing
  endogenous microbial {ROS} production.
\newblock {\em Nature Biotechnology}, 31:160--165, 2013.

\bibitem{allison_2011}
K.~R. Allison, M.~P. Brynildsen, and J.~J. Collins.
\newblock Metabolite-enabled eradication of bacterial persisters by
  aminoglycosides.
\newblock {\em Nature}, 473:216--220, 2011.

\bibitem{bollenbach_2015}
T.~Bollenbach.
\newblock Antimicrobial interactions: mechanisms and implications for drug
  discovery and resistance evolution.
\newblock {\em Curr. Opin. Microbiol.}, 27:1--9, 2015.

\bibitem{bollenbach_2009}
T.~Bollenbach, R.~Chait, and R.~Kishony.
\newblock Non-optimal microbial response to antibiotics underlies suppressive
  drug interactions.
\newblock {\em Cell}, 139:707--718, 2009.

\bibitem{lieberman_2014}
T.~D. Lieberman, K.~B. Flett, I.~Telin, T.~R. Martin, A.~J. Mc{A}dam, G.~P.
  Priebe, and R.~Kishony.
\newblock Genetic variation of a bacterial pathogen within individuals with
  cystic fibrosis provides a record of selective pressures.
\newblock {\em Nature Genetics}, 46:82--87, 2014.

\end{thebibliography}
\end{document}